\documentclass[journal, twoside, letter, final]{IEEEtran}

\usepackage{xcolor}
\usepackage{graphicx,epsfig,amssymb,amsmath,amsthm,amsfonts}
\usepackage{epstopdf}
\usepackage{multirow}
\usepackage{pict2e}
\usepackage{times}
\usepackage{cases}
\usepackage{array}
\usepackage{float}
\usepackage{algorithm}
\usepackage{algorithmic}
\usepackage{flushend}
\usepackage{cite}
\usepackage{bm}
\usepackage{setspace}
\usepackage{textcomp}
\usepackage{subfig}
\usepackage{enumerate}
\usepackage[font=small]{caption}

\begin{document}
\title{A Sparsity Adaptive Algorithm to Recover NB-IoT Signal from Legacy LTE Interference}
\author{Yijia~Guo, Wenkun~Wen,  Peiran~Wu, and Minghua~Xia,~\IEEEmembership{Senior~Member,~IEEE}%
\thanks{Manuscript received June 12, 2021; revised August 4, 2021; accepted September 10, 2021. This work was supported in part by the Key-Area Research and Development Program of Guangdong Province under Grant No. 2018B010114001, and in part by the National Natural Science Foundation of China under Grants 61801526, 62171486, and U2001213. The associate editor coordinating the review of this letter and approving it for publication was Kun Yang. \emph{(Corresponding author: Minghua Xia.)}}
\thanks{Y. Guo, P. Wu and M. Xia are with the School of Electronics and Information Technology, Sun Yat-sen University, Guangzhou 510006, China (e-mail: guoyj33@mail2.sysu.edu.cn, \{wupr3, xiamingh\}@mail.sysu.edu.cn).}
\thanks{W. Wen is with the Guangzhou Techphant Co. Ltd., Guangzhou 510310, China (e-mail: wenwenkun@techphant.net).}
}
\maketitle
\begin{abstract}
As a forerunner in 5G technologies, Narrowband Internet of Things (NB-IoT) will be inevitably coexisting with the legacy Long-Term Evolution (LTE) system. Thus, it is imperative for NB-IoT to mitigate LTE interference. By virtue of the strong temporal correlation of the NB-IoT signal, this letter develops a sparsity adaptive algorithm to recover the NB-IoT signal from legacy LTE interference, by combining $K$-means clustering and sparsity adaptive matching pursuit (SAMP). In particular, the support of the NB-IoT signal is first estimated coarsely by $K$-means clustering and SAMP algorithm without sparsity limitation. Then, the estimated support is refined by a repeat mechanism. Simulation results demonstrate the effectiveness of the developed algorithm in terms of recovery probability and bit error rate, compared with competing algorithms.
\end{abstract}
\IEEEpeerreviewmaketitle

\section{Introduction}
\IEEEPARstart{W}{ith} the online convening of the ITU-R WP5D \#35 meeting held on July 9th, 2020, Narrowband Internet of Things (NB-IoT) has been formally approved as part of the 5G standard, aiming at massive machine-type communications. As a forerunner in 5G ecosystem construction and industrial applications, NB-IoT will be inevitably coexisting with legacy Long-Term Evolution (LTE) systems. In real-world applications, NB-IoT can operate in three distinct modes: stand-alone, guard-band, or in-band. Among them, the in-band mode impacts NB-IoT and LTE performance as it bundles NB-IoT directly into the LTE's channels. Thus, eliminating interference from legacy LTE signal benefits the performance of in-band operating NB-IoT.

As the bandwidth of NB-IoT is much smaller than that of LTE, the NB-IoT signal is sparse in the frequency domain. Accordingly, recovering the NB-IoT signal from wideband LTE interference can be modeled as a sparse recovery problem. In recent years, the theory of compressed sensing (CS) \cite{Eldar12} was widely used to solve various sparse recovery problems in wireless systems. For instance, in \cite{Gomaa10} the CS theory was exploited to cancel narrow-band interference (NBI) in orthogonal frequency division multiplexing (OFDM) systems. In \cite{Al-Tous16, Al-Tous18}, several CS-based greedy algorithms were developed for joint data recovery and NBI mitigation in OFDM systems. In \cite{Zhang18}, the CS theory was applied to achieve both high detection and low false alarm probabilities for wideband spectrum sensing in cognitive radio systems. On the other hand, there have been extensive works on designing CS-based greedy algorithms, such as subspace pursuit \cite{Dai09}, sparsity adaptive matching pursuit (SAMP) \cite{Do08}, and iterative reweighted least-squares \cite{Xie21}. However, the CS theory requires that an observation matrix must satisfy the restricted isometry property \cite{Daubechies10}, which does not always hold in practice. 

Unlike the statistical schemes described above, machine learning was recently applied in \cite{Liu19} to eliminate NB-IoT interference to LTE system, where the support of NB-IoT subcarriers was located by an initial support distribution vector. In particular, this initial vector was first used to generate candidate supports from which several favorable ones were chosen, then the support distribution vector was updated as per the favorable ones by minimizing their cross-entropy. However, NB-IoT signal was simply modeled as an NBI in \cite{Liu19}, without exploiting the signal characteristics. By accounting for the continuous distribution of multiple NB-IoT subcarriers, a recent work \cite{Guo21} developed an intelligent recovery algorithm based on $K$-means clustering, where the support of subcarriers was first estimated based on $K$-means strategy and then located by a sliding window whose length equals the prior sparsity of NB-IoT signals.

For in-band operating NB-IoT, as the coherence time of NB-IoT signal is much larger than that of one OFDM symbol\cite{Liu17}, the NB-IoT signal is characterized by strong \emph{temporal correlation}, implying it has invariant support over one received OFDM frame of interest \cite{Liu19}. On account of the temporal correlation effect, this letter designs a sparsity adaptive recovery algorithm by combining $K$-means clustering and SAMP algorithm. In particular, the support of the NB-IoT signal is first estimated coarsely by $K$-means clustering and SAMP algorithm without sparsity limitation. Then, the estimated support is refined by a repeat mechanism. Compared with state-of-the-art competing algorithms, extensive simulation results demonstrate the effectiveness of the developed algorithm in terms of recovery probability, bit error rate, and convergence.

\textit{Notation}: Vectors and matrices are denoted by lower- and upper-case letters in bold typeface. The superscripts $(\cdot)^{\dagger}$ and $(\cdot)^H$ indicate the pseudo-inverse and conjugate transpose, respectively. The matrices $\bm{F}_N$ and $\bm{F}_N^H$ with size $N \times N$ refer to discrete Fourier transform (DFT) and inverse discrete Fourier transform (IDFT), respectively. The symbol $\tilde{\bm{x}}$ transforms $\bm{x}$ in time domain into frequency domain. The norms $\left\| \cdot \right\|_0$ and $\left\| \cdot \right\|_2$ stand for the $\ell_0$- and $\ell_2$-norm, respectively. The operator $\text{Card}\left( \mathcal{S}\right) $ gives the cardinality of set $\mathcal{S}$ and $\bm{W}_{\mathcal{S}}$ indicates the submatrix of $\bm{W}$, whose columns are indexed by set $\mathcal{S}$.

\section{System Model}
\label{Section-II}
\begin{figure}[H]
	\centering
	\includegraphics[width=2.55in, clip, keepaspectratio]{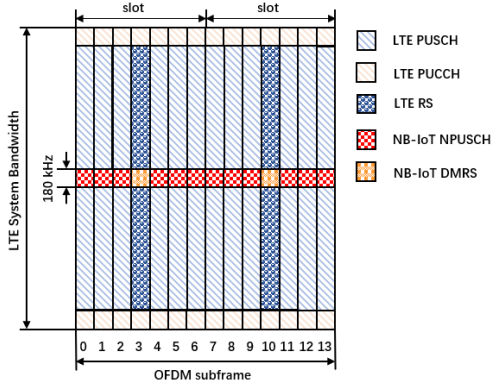}
	\caption{The time-frequency frame structure of in-band NB-IoT (PUSCH: physical uplink shared channel; PUCCH: physical uplink control channel; RS: reference signal; NPUSCH: narrowband physical uplink shared channel; DMRS: demodulation reference signal).}
	\label{Fig-1}
	\vspace{-8pt}
\end{figure}

As shown in Fig.~\ref{Fig-1}, the base station (BS) multiplexes the NB-IoT and LTE traffic onto the same frequency spectrum. To be specific, the NB-IoT occupies one physical resource block (PRB) in the LTE system bandwidth while the LTE occupies the other PRBs. Obviously, the mutual interference between NB-IoT and LTE signals is inevitable.

\begin{figure}[H]
	\centering
	\includegraphics[width=2.85in, clip, keepaspectratio]{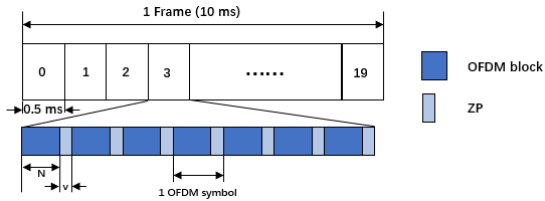}
	\caption{The ZP-OFDM frame structure of LTE.}
	\label{Fig-2}
	\vspace{-8pt}
\end{figure}

Figure~\ref{Fig-2} depicts the frame structure of LTE. It is clear that the zero-padding (ZP) is the guard interval between two consecutive OFDM blocks, so as to eliminate the inter-symbol interference (ISI). As a result, each OFDM symbol is composed of an OFDM block consisting of $N$ subcarriers and a length-$v$ zero sequence.

Suppose the channel between an LTE user equipment (UE) and its associated BS has a channel impulse response (CIR) $\bm{h} \triangleq \begin{bmatrix} h(0), h(1), \cdots, h(L) \end{bmatrix}$ where $L$ is the length of CIR, then, the received LTE signal at the BS can be formulated as 
\begin{equation} \label{Eq-1}
    \bm{y}_\text{LTE}=\bm{H}_\text{LTE}\bm{T}_\text{ZP}\tilde{\bm{x}}_\text{LTE},
\end{equation}
where $\tilde{\bm{x}}_\text{LTE}$ denotes the transmitted LTE signal in the frequency domain; $\bm{T}_\text{ZP} \triangleq \begin{bmatrix} \bm{F}_N & \bm{0}_{N \times v} \end{bmatrix}^H$ indicates the zero-padding process; $\bm{H}_\text{LTE}$ is modeled as a $P \times P$ lower triangular Toeplitz matrix whose first column is $\begin{bmatrix} \bm{h}&\bm{0}_{1\times (P-L-1)} \end{bmatrix}^H$, where $P \triangleq N+v$. In view of all-zero submatrix $\bm{0}_{v \times N}$ in $\bm{T}_{\text{ZP}}$, $\bm{H}_\text{LTE}$ can be expressed as a circulant matrix, with first row being $\begin{bmatrix} h(0), \bm{0}_{1\times (P-L-1)}, h(L),\cdots, h(1) \end{bmatrix}$. 

\begin{figure}[H]
	\centering
	\includegraphics[width=2.85in, clip, keepaspectratio]{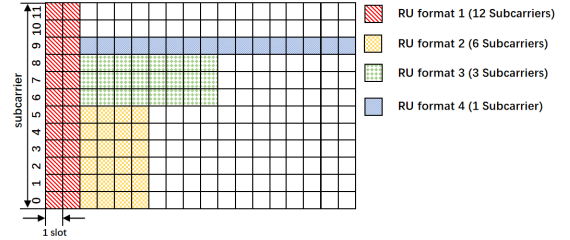}
	\caption{The deployment of different RUs in NPUSCH.}
	\label{Fig-3}
	\vspace{-8pt}
\end{figure}

As specified in Release 13 of the 3GPP specification TS 36.211 \cite{TS36211}, a transport data block of NB-IoT is first generated through channel coding, modulation and resource mapping, and then  transmitted over resource units (RUs). As illustrated in Fig.~\ref{Fig-3}, there are four distinct RU formats in NPUSCH, with each occupying different subcarriers and slots. For instance, the RU format 1 occupies $12$ subcarriers and $2$ slots whereas format 4 occupies $1$ subcarrier and $16$ slots.

Based on the aforementioned discussion, the received NB-IoT signal at the BS can be formulated as a sparse signal:
\begin{equation} \label{Eq-2}
    \tilde{\bm{y}}_\text{NB} =
    \begin{bmatrix}
        \tilde{y}_\text{NB}(0), \tilde{y}_\text{NB}(1), \cdots, \tilde{y}_\text{NB}(P-1)
    \end{bmatrix}^H,
\end{equation}
with
\begin{equation} \label{Eq-3}
    \tilde{y}_\text{NB}(t) = \left\{ 
    \begin{aligned}
        a_t, & \ m \leq t \leq n; \\
        0, & \ \text{otherwise},
    \end{aligned}
    \right.
\end{equation}
where $a_t \in \mathbb{C}$ is the signal amplitude at the $t^{\rm th}$ subcarrier, and $m \ge 0$ and $n \le P-1$ are the indices of the first and last subcarriers of the NB-IoT signal, respectively. $\mathcal{S}\triangleq\left\{ t|\tilde{y}_\text{NB}(t) \neq 0, \ t=0, 1, \cdots, P-1\right\}$ is the support of $\tilde{\bm{y}}_\text{NB}$, namely, the index set of nonzero elements of $\tilde{\bm{y}}_\text{NB}$. Also, it is evident that $\text{Card}\left( \mathcal{S}\right)  =  N_{\rm sc}$ with $N_{\rm sc}\in \left\lbrace 1, 3, 6, 12\right\rbrace$, as configured in Fig.~\ref{Fig-3}.

In light of \eqref{Eq-1} and \eqref{Eq-2}, the received signal in time domain at the BS can be expressed as
\begin{align}
    \bm{y} &= \bm{y}_\text{LTE}+\bm{F}_P^{H}\tilde{\bm{y}}_\text{NB}+\bm{n} \label{Eq-4} \\
    &= \bm{H}_\text{LTE}\bm{T}_\text{ZP}\tilde{\bm{x}}_\text{LTE}+\bm{F}_P^{H}\tilde{\bm{y}}_\text{NB}+\bm{n}, \label{Eq-5}
\end{align}
where $\bm{n}$ denotes an additive white Gaussian noise (AWGN) with zero mean and covariance matrix $\sigma^2 \bm{I}$. Moreover, the instantaneous received power of the NB-IoT signal and LTE signal can be explicitly computed as
\begin{equation} \label{Eq-6}
    P_{\text{NB}} = \frac{\bm{y}_\text{NB}^H\bm{y}_\text{NB}}{N_{\rm sc}}, \quad
    P_{\text{LTE}} = \frac{\bm{y}_\text{LTE}^H\bm{y}_\text{LTE}}{N},
\end{equation}
respectively. By definition, the received signal-to-noise ratio (SNR) is given by $P_{\text{NB}}/\sigma^2$ while the received signal-to-interference ratio (SIR) is $P_{\text{NB}}/P_{\text{LTE}}$.

\section{Algorithm Development}
\label{Section-III}
In this section, we first preprocess the received signal given by \eqref{Eq-5} so as to eliminate the interference caused by the LTE signal. Then, a sparsity adaptive algorithm combining $K$-means clustering and SAMP algorithm is developed to efficiently recover NB-IoT signal.

\subsection{Preprocessing the Received Signal}
\label{Section-III-A}
As $\bm{H}_\text{LTE}$ defined after \eqref{Eq-1} is a circulant matrix, it can be diagonalized such that \eqref{Eq-5} can be rewritten as 
\begin{align}
     \bm{y}  = \bm{F}_P^{H}\bm{\Lambda}_\text{LTE}\bm{F}_P\bm{T}_\text{ZP}\tilde{\bm{x}}_\text{LTE}+\bm{F}_P^{H}\tilde{\bm{y}}_\text{NB}+\bm{n}, \label{Eq-7}
\end{align}
where $\bm{\Lambda}_\text{LTE}$ is a diagonal matrix. After performing DFT over \eqref{Eq-7}, we obtain the received signal in frequency domain:
\begin{equation} \label{Eq-8}
    \tilde{\bm{y}} = \bm{\Lambda}_\text{LTE}\bm{F}_P\bm{T}_\text{ZP}\tilde{\bm{x}}_\text{LTE} + \tilde{\bm{y}}_\text{NB} + \tilde{\bm{n}}. 
\end{equation}

Now, we are in a position to mitigate the LTE signal in \eqref{Eq-8}. In light of $\bm{F}_P$ being an orthogonal matrix and $\bm{\Lambda}_\text{LTE}$ being a diagonal matrix, let $\bm{G}_1\triangleq\bm{F}_P^H\bm{\Lambda}_\text{LTE}^{-1}$, then, multiplying $\tilde{\bm{y}}$ by $\bm{G}_1$ gives 
\begin{equation} \label{Eq-9}
    \bm{y}_1 = \bm{T}_\text{ZP}\tilde{\bm{x}}_\text{LTE} + \bm{G}_1\tilde{\bm{y}}_\text{NB} + \bm{G}_1\tilde{\bm{n}}.
\end{equation}
Next, by recalling the special structure of the ZP-OFDM system illustrated in Fig.~\ref{Fig-2}, we can fully eliminate the LTE signal by using the zero-padding appended at the end of each OFDM block. To be specific, let $\bm{G}_2 \triangleq \begin{bmatrix} \bm{0}_{v \times N} & \bm{I}_v \end{bmatrix}$, then, multiplying $\bm{y}_1$ by $\bm{G}_2$ yields 
\begin{align}
    \bm{y}_2 = \bm{G}_2\bm{G}_1\tilde{\bm{y}}_\text{NB}+\bm{G}_2\bm{G}_1\tilde{\bm{n}} 
     = \bm{W}\tilde{\bm{y}}_\text{NB}+\bm{W}\tilde{\bm{n}}, \label{Eq-10}
\end{align}
where $\bm{W} \triangleq \bm{G}_2\bm{G}_1 \in \mathbb{C}^{v \times P}$ is known as an observation matrix. Thus, our remaining task is to recover the sparse signal $\tilde{\bm{y}}_\text{NB}$ from the postprocessed $\bm{y}_2$. 

As shown in Fig.~\ref{Fig-3}, when NB-IoT operates in the in-band mode, the NB-IoT signal in frequency domain occupies several consecutive subcarriers. More accurately, the number of subcarriers is allowed to be $1$, $3$, $6$ or $12$, i.e., $N_{\rm sc}\in \left\lbrace 1, 3, 6, 12\right\rbrace$. This feature implies that $\bm{y}_2$ shown in \eqref{Eq-10} is a linear combination of several consecutive column vectors of $\bm{W}$. In other words, the correlation coefficients of $\bm{y}_2$ and each column vector of $\bm{W}$ will show one or more spikes, which can be located by the $K$-means clustering.

Applying the least squares (LS) principle to \eqref{Eq-10}, the optimization problem can be formulated as
\begin{align}
    \mathcal{P}1:\ &\tilde{\bm{y}}_\text{NB}^{*} = \underset{\tilde{\bm{y}}_\text{NB}}{\arg \min }\left\|\bm{y}_2-\bm{W} \tilde{\bm{y}}_\text{NB}\right\|_{2},  \label{Eq-11} \\
    &\text { s.t. }\left\|\tilde{\bm{y}}_\text{NB}\right\|_{0} \in \left\lbrace 1, 3, 6, 12\right\rbrace, \label{Eq-12}
\end{align}
where the objective function is indeed the $\ell_2$-norm of the residue error, that is, $r \triangleq \left\|\bm{y}_2-\bm{W} \tilde{\bm{y}}_\text{NB}\right\|_{2}$.

\subsection{Coarse Estimation of the Support}
\label{Section-III-B}
In principle, finding the support of $\tilde{\bm{y}}_\text{NB}$ amounts to choosing a set of consecutive column vectors of $\bm{W}$ such that their linear combination minimizes the residue error in \eqref{Eq-11}. To this end, the correlation coefficient of $\bm{y}_2$ and each column of $\bm{W}$ is computed as
\begin{equation} \label{Eq-13}
	{\gamma}(i)=\frac{\left| {\bm{y}_2^H\bm{W}(:,i)}\right| }{\left\| \bm{W}(:,i)\right\| _2}, \ \forall i=1, \cdots, P.
\end{equation} 

On account of the continuous distribution of the support of $\tilde{\bm{y}}_\text{NB}$ (cf. Fig.~\ref{Fig-3}), we employ the $K$-means clustering algorithm to classify these coefficients by Euclidean distance \cite{Bishop06}. Suppose that we obtain $Q$ clusters, say, $\left\{ \mathcal{C}_q\right\} _{q=1}^Q$, where $\mathcal{C}_q$ contains correlation coefficients in the $q^{\rm th}$ cluster, then, the optimal cluster is determined by 
\begin{equation} \label{Eq-14}
	\mathcal{C}_{\text{opt}} = \arg \max \left\{\frac{\sum \gamma(i)}{\text{Card}\left( \mathcal{C}_q\right) } \right\}_{q=1}^{Q}, \ \forall \gamma(i)\in\mathcal{C}_q,
\end{equation}
where $\mathcal{C}_{\text{opt}}$ is the cluster with the largest mean value of the correlation coefficients. As a result, the optimal column index set is given by 
\begin{equation} \label{Eq-15}
	\mathcal{S}_{\text{opt}} = \left\{ i\left| \gamma(i)\in \mathcal{C}_{\text{opt}}\right. \right\}.  
\end{equation}

Now, since $\bm{y}_2$ is more likely a linear combination of column vectors whose column indices belong to $\mathcal{S}_{\text{opt}}$, the optimization problem $\mathcal{P}1$ can be further simplified as
\begin{align}
    \mathcal{P}2:\ &\tilde{\bm{y}}_\text{NB}^{*}=\underset{\tilde{\bm{y}}_\text{NB}}{\arg \min }\left\|\bm{y}_2-\bm{W}_{\mathcal{S}_{\text{opt}}} \left( \tilde{\bm{y}}_\text{NB}\right)_{\mathcal{S}_{\text{opt}}} \right\|_{2}, \label{Eq-16} \\
    &\text { s.t. }\left\|\left( \tilde{\bm{y}}_\text{NB}\right)_{\mathcal{S}_{\text{opt}}}\right\|_{0} \in \left\lbrace 1, 3, 6, 12\right\rbrace. \label{Eq-17}
\end{align}
To deal with $\mathcal{P}2$, we define a count vector $\bm{f}_{P \times 1}$ initialized as a null vector, then, the SAMP algorithm is used to solve the minimization problem in \eqref{Eq-16} without considering the sparsity constraint in \eqref{Eq-17}. With the estimated support denoted $\mathcal{I}$, the count vector $\bm{f}$ is updated as per 
\begin{equation} \label{Eq-18}
    \bm{f}(j) \gets \bm{f}(j)+1, \ \forall j \in \mathcal{I}.
\end{equation}

It is noteworthy that, in practice, the column vectors chosen by $\mathcal{S}_{\text{opt}}$ shown in \eqref{Eq-15} may be linearly dependent. If so, this dependence would degrade the recovery probability. To deal with this situation, we randomly disturb the columns of $\bm{W}$ to form a new observation matrix $\bm{W}^{\prime}$ and corresponding measurement vector $\bm{y}_2^{\prime}$. According to \eqref{Eq-13}-\eqref{Eq-15}, a new optimal column index set $\mathcal{S}_{\text{opt}}^{\prime}$ is generated and $\bm{W}_{\mathcal{S}_{\text{opt}}}$ in $\mathcal{P}2$ is replaced by $\bm{W}_{\mathcal{S}_{\text{opt}}^{\prime}}$. Then, the SAMP algorithm is repeatedly used to solve the problem in \eqref{Eq-16} again. The estimated support is denoted $\mathcal{I}^{\prime}$, and the count vector $\bm{f}$ proceeds with updating by $\mathcal{I}^{\prime}$ according to \eqref{Eq-18}. The aforementioned operations are repeated $R_{\max}$ times to obtain the final count vector $\bm{f}$, which is next used to refine the estimate of the support.

\subsection{Refined Estimation of the Support}
\label{Section-III-C}
To minimize the $\ell_2$-norm of the residue error in \eqref{Eq-16}, the estimated support $\mathcal{I}$ by the SAMP at each repetition described above does not satisfy the constraint in \eqref{Eq-17}. Now, we exploit the final count vector $\bm{f}$ to refine the estimate of the support. It is not hard to understand that, after $R_{\max}$ repetitions, the final $\bm{f}$ contains the distribution information of the real support of NB-IoT signal. Since NB-IoT occupies several continuous subcarriers in the LTE spectrum, finding the real support of the NB-IoT signal is equivalent to finding an index range in $\bm{f}$ where an spike occurs.

The procedure to recover the support of NB-IoT signal from the final count vector $\bm{f}$ is as follows. First, we define a difference vector $\bm{d} \in \mathbb{R}^{P \times 1}$ with entries given by 
\begin{equation} \label{Eq-19}
	\bm{d}(l) = \left\{
		\begin{array}{rl}
			\bm{f}(l)-\bm{f}(l+1), & \text{if } 1\leq l < P; \\
			\bm{f}(P)-\bm{f}(1), & \text{if } l = P.
		\end{array} \right. 
\end{equation}
Then, given the indices of the minimum and maximum values in the difference vector $\bm{d}$, denoted $p_1$ and $p_2$, respectively, determined by
\begin{align}
	p_1 =\arg\min\limits_{l = 1, \cdots, P}\left\lbrace \bm{d}(l)\right\rbrace,  \quad 
	p_2 =\arg\max\limits_{l = 1, \cdots, P}\left\lbrace \bm{d}(l)\right\rbrace, \label{Eq-20}
\end{align}
the starting point and endpoint of the spike are located. Next, with $p_1$ and $p_2$, a sample set $\mathcal{X}$ can be generated as per
\begin{equation} \label{Eq-21}
\mathcal{X}=\left\{
\begin{array}{rl}
\left\lbrace \bm{f}(p_1+1), \cdots, \bm{f}(p_2)\right\rbrace, & \text{if } 1 \leq p_1 < P; \\
\left\lbrace \bm{f}(1), \cdots, \bm{f}(p_2)\right\rbrace,  & \text{if } p_1 = P.
\end{array} \right.
\end{equation}
If ${\rm var}(\mathcal{X})<\epsilon$, then it implies that the elements of $\bm{f}$ between indices $p_1$ and $p_2$ shape one spike. Further, if $\text{Card}\left( \mathcal{X}\right)  \in \left\lbrace 1, 3, 6, 12\right\rbrace $, the estimated optimal support is finally given by
\begin{equation} \label{Eq-22}
\mathcal{Z}^{*} = \left\{
\begin{array}{rl}
(p_1+1:p_2), & \text{if } 1\leq p_1 < P; \\
(1:p_2),  & \text{if } p_1 = P.
\end{array} \right.
\end{equation}
Otherwise, repeat the aforementioned operations until ${\rm var}(\mathcal{X})<\epsilon$ and $\text{Card}\left( \mathcal{X}\right)  \in \left\lbrace 1, 3, 6, 12\right\rbrace $. Simulation results to be discussed at the end of the next section demonstrate that the algorithm converges after $30$ repetitions. 

With the optimal support estimated, the transmitted NB-IoT signal can be recovered by using the LS principle. Specifically, let $\bm{y}_\text{LS} \triangleq \bm{W}_{\mathcal{Z}^{*}}^{\dagger}\bm{y}_2$, then $\tilde{\bm{y}}_\text{NB}^{*}$ can be expressed as
\begin{equation} \label{Eq-23}
	\tilde{\bm{y}}_\text{NB}^{*}(t) = \left\{ 
	\begin{array}{rl}
		\bm{y}_\text{LS}\left(t-\mathcal{Z}^{*}(1)+1\right), & \text{if } t \in \mathcal{Z}^{*}; \\
	0, & \text{otherwise}.
\end{array} \right. 
\end{equation}

To sum up, the developed algorithm to recover the NB-IoT signal from LTE interference is formalized in Algorithm~\ref{Algorithm-1}.

\section{Simulation Results and Discussions}
\label{Section-IV}
In this section, we present and discuss the computational complexity and simulation results pertaining to the developed algorithm. In the simulation experiments, to generate the NB-IoT signal, the information bits are encoded by Turbo code with code rate $1/3$ and the coded bits are modulated with quadrature phase shift keying (QPSK) constellation. As for the LTE signal, the OFDM block length is set to $N = 600$ with ZP length $v = 144$. The length of CIR is $L = 50$. In the signal recovery algorithm, the maximum iteration number is set to $I_{\max}=30$ with each the maximum repetition number $R_{\max}=50$, and the number of clusters generated by $K$-means clustering is $Q=30$. The number of NB-IoT subcarriers is specified as $\{1, 3, 6, 12\}$, which coincides with the real configuration shown in Fig.~\ref{Fig-3}.

\begin{algorithm} [!t] 
	\caption{Recovering NB-IoT Signal for LTE Interference} 
	\label{Algorithm-1}  
	\small
	\begin{algorithmic}[1] 
		\REQUIRE The measurement vector $\bm{y}_2^{(1)}=\bm{y}_2$, the observation matrix $\bm{W}^{(1)} \triangleq \bm{W}$, the maximum iteration number $I_{\max}$, the maximum repetition number $R_{\max}$, the variance threshold $\epsilon$, and the received signal $\tilde{\bm{y}}$;
		\ENSURE The estimated support $\mathcal{Z}^{*}$ and NB-IoT signal $\tilde{\bm{y}}_\text{NB}^{*}$;
		\STATE {\bf Initialization:} $\bm{f}=\bm{0}_{P\times 1}$, $t=1$, $\mathcal{A}^{(1)} = \emptyset$;
		\WHILE{$t\leq I_{\max}$} 
		\FOR{$k=1:R_{\max}$} 
		\FOR{$i=1:P$} 
		\STATE Calculate $\gamma(i)$ according to \eqref{Eq-13};
		\STATE Add $\gamma(i)$ to the end of $\mathcal{A}^{(k)}$;
		\ENDFOR 
		\STATE Apply the $K$-means clustering algorithm to $\mathcal{A}^{(k)}$;
		\STATE Compute $\mathcal{C}_{\text{opt}}$ and $\mathcal{S}_{\text{opt}}$ as per \eqref{Eq-14} and \eqref{Eq-15}, resp.;
		\STATE Recover $\mathcal{I}$ by solving \eqref{Eq-16} with SAMP algorithm;
		\STATE Update the count vector $\bm{f}$ according to \eqref{Eq-18};
		\STATE Update the columns in $\bm{W}^{(k)}$ to form $\bm{W}^{(k+1)}$;
		\STATE $\bm{y}_2^{(k+1)} = \bm{W}^{(k+1)}\tilde{\bm{y}}$;
		\STATE $\mathcal{A}^{(k+1)} = \emptyset$;
		\ENDFOR
		\STATE Compute the difference vector $\bm{d}$ according to \eqref{Eq-19};
		\STATE Find $p_1$ and $p_2$ as per \eqref{Eq-20}, and generate sample set $\mathcal{X}$ by \eqref{Eq-21};
		\IF{$\text{Card}\left( \mathcal{X}\right)  \in \left\lbrace 1, 3, 6, 12\right\rbrace $ and ${\rm var}(\mathcal{X}) \leq \epsilon$} 
		\STATE Estimate the support $\mathcal{Z}^{*}$ according to \eqref{Eq-22};
		\STATE break;
		\ENDIF
		\STATE $\bm{f}=\bm{0}_{P \times 1}$;
		\STATE $t=t+1$;
		\ENDWHILE 
		\STATE Compute $\tilde{\bm{y}}_\text{NB}^{*}$ according to \eqref{Eq-23};
	\end{algorithmic} 
\end{algorithm}

\subsection{Computational Complexity Analysis}
\label{Section-IV-A}
Table~\ref{table-1} compares the computational complexity of the proposed algorithm with three benchmark ones developed in \cite{Do08, Liu19, Guo21}. It is observed that the proposed algorithm has lower complexity than the classic SAMP \cite{Do08}, as the latter locates directly the support of NB-IoT subcarriers without any pre-estimate. The computational complexity of our algorithm and that in \cite{Guo21} (denoted `CWS' for short) is independent of the number of NB-IoT subcarriers (i.e., $K$), whereas that in \cite{Liu19} (denoted `SCEM' for short) increases quadratically with $K$. The proposed algorithm has a bit higher complexity than that in \cite{Guo21} because the latter assumes the prior sparsity of NB-IoT signal, which is however unknown in practice.
\begin{table}[!t]
\centering
\renewcommand{\arraystretch}{1.1}
\caption{Comparison of Algorithm Complexity}
\label{table-1}
\small
\begin{tabular}{cc}
\hline
Algorithm & Computational Complexity\\
\hline
SAMP \cite{Do08} & $\mathcal{O}(R_{\max}v^2P^2)$\\
\hline
SCEM \cite{Liu19} & $\mathcal{O}(I_{\max}R_{\max}(N_{c}vK^2+N_{f}P))$\\
\hline
CWS \cite{Guo21} & $\mathcal{O}(R_{\max}P^2)$\\
\hline
Proposed & $\mathcal{O}(I_{\max}R_{\max}P^2)$\\
\hline
\end{tabular}
\vspace{-15pt}
\end{table}

\subsection{Simulation Results and Discussions}
\label{Section-IV-B}
\begin{figure}[!t]
	\centering
	\includegraphics[width=2.5in, clip, keepaspectratio]{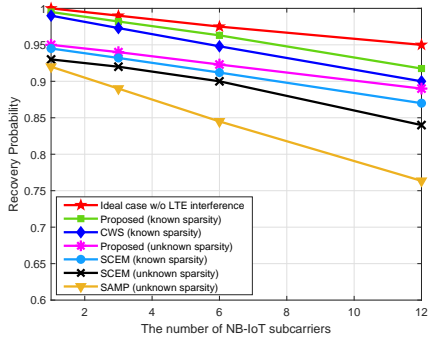}
	\caption{The recovery probability vs. the number of subcarriers.}
	\label{Fig-4}
	\vspace{-5pt}
\end{figure}

Figure~\ref{Fig-4} shows the recovery probability versus the number of NB-IoT subcarriers. In the pertaining simulation setup, the SNR is fixed to $23$ dB while the SIR is $20$ dB, which means that the system is interference dominated. It is seen from Fig.~\ref{Fig-4} that the recovery probability decreases with the number of NB-IoT subcarriers, as more subcarriers suffer higher interference. With sparsity known, the developed algorithm outperforms the SCEM and CWS algorithm and approaches the performance in the ideal case. On the other hand, in the case of unknown sparsity, the developed algorithm performs much better than the SCEM and SAMP algorithm.

\begin{figure}[!t]
	\centering
	\includegraphics[width=2.5in, clip, keepaspectratio]{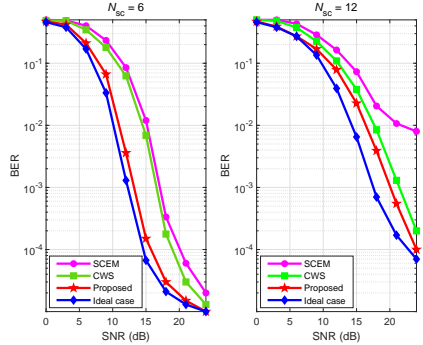}
	\caption{The BER vs. the SNR of NB-IoT signal with {\it known sparsity}.}
	\label{Fig-5}
	\vspace{-5pt}
\end{figure}

Figure~\ref{Fig-5} illustrates the bit error rate (BER) versus the SNR of the NB-IoT signal with {\it known sparsity}, where the SIR is fixed to $15$ dB and $N_{\rm sc} = 6$ in the left panel (or $N_{\rm sc} = 12$ in the right panel). It is seen that the BER decreases with SNR but increases with the number of NB-IoT subcarriers, as expected. In the case of $N_{\rm sc} = 6$, given the target BER being $10^{-3}$, the left panel of Fig.~\ref{Fig-5} shows that the required SNR corresponding to the ideal case is about $12$ dB and the SNR required by the proposed algorithm is about $13$ dB, which outperforms the CWS algorithm by $3.5$ dB and the SCEM algorithm by $4$ dB.

\begin{figure}[!t]
	\centering
	\includegraphics[width=2.5in, clip, keepaspectratio]{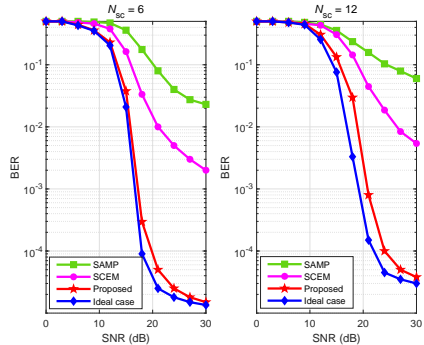}
	\caption{The BER vs. the SIR of NB-IoT signal with {\it unknown sparsity}.}
	\label{Fig-6}
	\vspace{-5pt}
\end{figure}

Figure~\ref{Fig-6} depicts the BER versus the SNR of the NB-IoT signal with  {\it unknown sparsity}, where the SIR is fixed to $20$ dB and $N_{\rm sc} = 6$ in the left panel (or $N_{\rm sc} = 12$ in the right panel). It shows that the BER of the proposed algorithm approaches that in the ideal case without interference whereas the SAMP and SCEM algorithm perform much worse. 

\begin{figure}[!t]
	\centering
	\includegraphics[width=2.5in, clip, keepaspectratio]{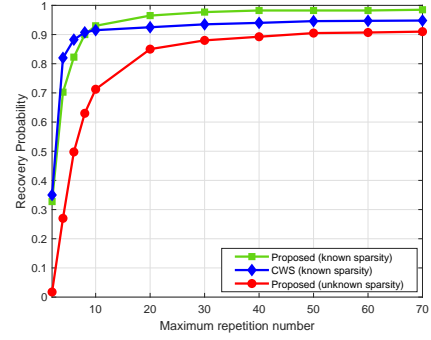}
	\caption{The recovery probability vs. the maximum repetition number.}
	\label{Fig-7}
	\vspace{-5pt}
\end{figure}

Figure~\ref{Fig-7} shows the recovery probability versus the maximum repetition number $R_{\max}$ in each iteration. In the simulation setup, the number of NB-IoT subcarriers is set to $N_{\rm sc} = 6$ with SNR being $23$ dB and SIR being $20$ dB. It is observed that all three curves converge after 30 repetitions. In the case of known sparsity, the proposed algorithm obtains $98\%$ recovery probability, which outperforms the CWS algorithm by $4\%$, albeit a bit slower convergence. In the case of unknown sparsity, the CWS algorithm does not work anymore but the proposed algorithm has a recovery probability about $90\%$.

\section{Conclusions}
\label{Section-V}
In this letter, a sparsity adaptive algorithm was designed to recover NB-IoT signal from legacy LTE interference. In particular, by using the strong temporal correlation of NB-IoT signal, the support of the NB-IoT signal was first estimated in a coarse way, and then was refined by a repeat mechanism. As the developed two-stage algorithm needs neither the restricted isometry property on the observation matrix nor the sparsity information of the NB-IoT signal, it is promising in real-world applications. 

\bibliographystyle{IEEEtran}
\bibliography{References}

\vfill
\end{document}